\documentclass{article}
\usepackage[T1]{fontenc}
\usepackage[utf8]{inputenc}
\usepackage{textcomp}
\usepackage{eurosym}

\usepackage[textsize=footnotesize]{todonotes}

\usepackage{hyperref}
\hypersetup{
	colorlinks = true,
    linkcolor = blue,
    urlcolor  = blue,
    citecolor = blue,
    anchorcolor = blue]
}
\usepackage[backend=bibtex,style=numeric-comp,natbib=true,backref=true,url=false,doi=false]{biblatex}
\usepackage[T1]{fontenc}
\usepackage[margin=3.5cm]{geometry}
\usepackage{authblk}
\usepackage{graphicx}

\usepackage{scalerel} % a package for scaling inline images as compared to text size
\usepackage{float}
\usepackage[labelfont=bf,font=small,margin=0.5cm]{caption}
\usepackage[font=scriptsize]{subcaption}
\usepackage{mdframed}

\usepackage[printwatermark]{xwatermark}
\usepackage{xcolor}
\newwatermark[allpages,color=lightgray!50,angle=45,scale=3,xpos=0,ypos=0]{DRAFT}

\usepackage{enumitem}

\setlist[enumerate,1]{%
  label=\roman*.,
}
  
\newlist{inlinelist}{enumerate*}{1}
\setlist*[inlinelist,1]{%
  label=(\arabic*),
}

\usepackage{pdfpages}

\title{Policy Scan and Technology Strategy Design methodology}
\author[1,2]{Viktoras Kabir Veitas%
  \thanks{Electronic address: \texttt{vveitas@gmail.com}; Corresponding author}}
\author[1]{Simon Delaere}
\affil[1]{imec-SMIT Vrije Universiteit Brussel}
\affil[2]{Center Leo Apostel - VUB}
\affil[3]{Evolution, Complexity \& Cognition Group - VUB}
\date{February 15, 2018}

\begin{document}

\maketitle
\begin{abstract}

Increasingly accelerating technology advancement affects and disrupts almost all aspects of human society and civilization at large as we know it \citep{veitas_living_2017}. Actually, this has been true since the technology started at the dawn of human society, yet the mere speed and magnitude of modern technology development brings about the situation where societies and economies have to adapt to the changing technological landscape as much as technologies have to integrate into the social fabric. The only way to to achieve such integration in a changing and unpredictable world is to enable and support the close interaction between the world of societal problems and expectations and the world of technology. Policy Scan and Technology Strategy design methodology presented in this paper was developed precisely for the purpose of addressing specific types of 'ill-defined' problems in terms of observing, analyzing and integrating technology developments and availabilities with policy requirements, social governance and societal expectations. The methodology consists of conceptual tools and methods for developing concrete actions and products for guiding technology adoption for social change (a.k.a. \textit{empowerment by design}). The method developed in this work is geared towards increasingly complex and uncertain situations where existing analysis and problem solving methods often fail due to many non-linearities inherent in the social and technology worlds and, especially, at their area of their inter intersection. The development of the methodology followed the grounded theory construction process which requires a close relation to a specific context of an application domain, determined by actual interaction between the worlds of societal problems and technology. The chosen application domain of this research is the intersection of smart mobility problematics and opportunities, the rising autonomous driving technology, data privacy, provenance and security challenges, policies and legislation. This paper is first of two in the series, explaining the methodology with the necessary reference to examples from the application domain. The second paper of the series details the context itself and the concrete technological solution that mitigates identified concrete societal problem of the chosen application domain.

\end{abstract}

\section{Introduction -- research questions and problematic}\label{research_questions}

Socially succesfull innovations are able to integrate the "application pull" determined by often implicit social needs with the "technology push" comming from the increasingly difficult to predict in advance technological breaktrhoughts. More generally, the challenge of integrating large scale technological advancements into social life poses a fundamental problem of how two large and fluid domains of human society can be made to negotiate their interactions in the most beneficial way. These domains are:
\begin{enumerate}
\item The world of social expectations and governance -- rules, regulations, policies, standards, institutional structures, legislative systems, perceived social expectations and problems;
\item The world of technology -- directions and areas of technology developments (e.g. genetics, biotechnology, artificial intelligence, material science, etc.) which empower individual humans and societies to affect themselves and the world around them in fundamentally new and more powerful ways -- bringing disruption to the established order.
\end{enumerate}

The goal of the present research was to formulate and test in the real setting a systematic methodology that would facilitate the dialogue between these two worlds allowing to formulate and drive the long term innovation roadmap as well as identify and implement sucessful individual innovations. In order to address this challenge we have formulated following research questions:
\begin{description}
  \item [Research Question RQ1:] How to ensure a smooth integration of disruptive technology into the social fabric (i.e. govern the change)?
  \item [Sub-question SQ1:] How to ensure that technology complies to policy requirements, directions and social expectations;
  \item [Sub-question SQ2:] How to ensure that policy requirements, directions and social expectations are aligned with disruptive changes brought by technology.
\end{description}

First thing to notice is that SQ1 and SQ2 are circularly related and therefore cannot be tacked alone or even in sequential order. The research sub-questions and, consecutively, the main research question RQ1 can be answered only by observing and facilitating the actual interaction between technology, social governance and the social construction process that this interaction entails.

Therefore, the challenge that we are set to address is in the domain of what is called "wicked", "ill-defined" or "ill-structured" problems -- problems that are hard to solve because of the incomplete, contradictory and changing requirements that often are difficult to recognize \citep{banathy_designing_1996,veitas_system_2007}. Moreover, since multiple requirements of a wicked problem are interrelated (as in the case of sub-questions SQ1 and SQ2), trying to analytically solve one problem without simultaneously addressing the other may only create confusion and additional problems. "Ill-defined" problems are heavily context dependent and cannot be even started to be addressed without considering a very specific context where they appear. In case of integrating technology and social governance systems, such specific context is a concrete technology or solution around which the interaction of social governance and technology development domains can be puzzled out.

This brings about the methodological principle for establishing the actionable interaction between the world of societal problems and expectations and the world of technology in terms of \textbf{mapping concrete societal problems that can be addressed with specific technological availabilities}. The principle is at the core of Policy Scan and Technology Strategy Design -- the methodology that resulted from this research. It allows to create a funnel for consolidating general and abstract ideas, developments and tendencies into concrete expectations, requirements and problems to be addressed by concrete technology solutions. The result of the process is an actionable technology strategy with concrete proposals for solutions which address specific social needs and problems. Policy Scan and Technology Strategy design is therefore a design inquiry aimed at conceiving and designing specific solutions in the specific context which are identified during the process of applying the methodology. The development of the methodology is therefore best understood as a process of building a grounded theory. 

All results of the research are documented in two papers -- methodological and application-specific. In the current paper we explain the methodology, its theoretical influences and show how it is applied for identifying and motivating concrete junction points between the worlds of societal problems and technology. In the second paper \citep{veitas_-vehicle_2018} the identified societal problem, the mitigating technical solution (Data Storage Device for autonomous vehicles) and the surrounding context are presented and explained. Note, that since the methodology requires close relation to a specific context, a mention of a specific technology in the description of the method is at times necessary, even though we attempted to separate the methodology and its application specifics as much as didactically possible. 

The contents of the current paper is the following. Section \ref{conceptual-framework} introduces the broad conceptual basis and influences. Section \ref{policy-scan-methodology} explains the methodology in general terms with a minimal relation to the application specific use case. Section \ref{identifying-junction-points} shows how the methodology is being applied for constructing a funnel for identifying concrete junction points between the world of societal problems and the world of technology.

\section{Conceptual framework}\label{conceptual-framework}

The conceptual framework of the methodology combines five theoretical domains: 
\begin{inlinelist}
\item Social system design, 
\item Grounded theory, 
\item Actor-Network Theory, 
\item the concept of the boundary object with the surrounding discussion and 
\item change methodology for complex social systems. 
\end{inlinelist}

\subsection{Social system design}\label{social-system-design}

Social system design \citep{banathy_designing_1996} is an intellectual technology of future-creating disciplined inquiry that realizes people's vision of the future society, their own expectations and expectations of their environment. It seeks to understand a problem situation as a system of interconnected, interdependent and interacting issues and use this understanding to create a design as a network of internally consistent solution ideas. As an extension of systems thinking and system dynamics domains, the social system design technology provides intellectual tools for dealing with feedback loops, non-linearity, self-organization and emergent behaviour, characteristic to complex adaptive systems. Such systems emerge from interaction of many heterogeneous agents where their collective behaviour cannot be trivially inferred from understanding individual properties of agents. Furthermore, in social systems, the nature of interaction between heterogeneous agents is due to their perspectives towards each other and towards the context of a situation where the interaction takes place \citep{weinbaum_world_2014}. Social system design empowers an owner of a design inquiry (the actor in a social system which initiates and coordinates the process) to enable achievement of its goals and strategies that take into account vested interests of all agents and their groups of an ecosystem via achieving non-trivial synergies and cooperation and therefore beneficial for the whole social system.

The process of design leads the designer and the system from an existing stage to a desired future via four major processes performed iteratively:
\begin{inlinelist}
  \item \textit{transcending} the state of the existing system and leaving it behind;
  \item \textit{envisioning} an image of the system that we wish to create;
  \item \textit{designing the system}, which, when implemented, transforms the existing state to the desired future state;
  \item \textit{presenting/displaying} the model(s) of the designed system \citep[p. 61]{banathy_designing_1996}.
\end{inlinelist} 
In this research we employ the intellectual technology of social system design (mostly via previous work of co-authors on change management of complex social systems -- see \ref{change-management-in-complex-systems}) for integrating expectations towards and availabilities of the developing technologies in order to guide the change. 

\subsection{Grounded theory}\label{grounded-theory}

Grounded theory \citep{glaser_discovery_2000} is a methodology aimed at discovery of theory from systematically obtaining and analysing data in social research. The theory was developed by sociologists B. Glaser and A. Strauss in the process of research that culminated with 1965 \textit{Awereness of Dying} book \citep{glaser_awareness_2017}. It is a general method of specific situation- embedded comparative analysis and is a way of arriving at a theory suited to its supposed uses with relation to that situation. A crucial yet somewhat subtle principle implicit in grounded theory is that it is itself \textit{a process} -- an ever- developing entity and never a finished product -- where assumptions and principles guiding the development of the theory individuate when applying it. Grounded theory is often contrasted to theory generated by logical deduction from a priori assumptions. Note, however, that grounded theory does not negate logical deduction but actually uses it to direct theoretical sampling further in the process -- yet theory is never assumed prior to the data analysis. This explicitly links the theory with data from which it is being generated and therefore implies that the results obtained by applying a theory to data cannot be completely separated from the process of theory construction -- in contrast to the more established hypothetico-deductive approach and falsifiability principle in science. This means that a theory or method has to be revisited and adapted at each instance of its use. Policy Scan and Technology Strategy design process realizes this principle by requiring to revisit and, if applicable, adapt the methodology for each environmental situation in which it is being applied.

\subsection{Actor-Network Theory}

Actor Network Theory (\textit{ANT}), which emerged during 1980s from the work of mainly Bruno Latour, Michael Callon and John Law, is a \textit{conceptual frame for exploring collective sociotechnical processes, whose spokespersons have paid particular attention to science and technologic activity} \citep{ritzer_actor_2005}. It asserts that science is a process of heterogeneous engineering in which the social, technical, conceptual and textual are puzzled together and transformed. Independent of their nature, all entities in a puzzle achieve significance only via relation to others, without presupposition of any ontology prior to that. Therefore, the 'volitional actor' in \textit{ANT} (\textit{actant}) is any agent -- social, technical, human, collective or individual -- that can associate or dissociate with other agents by affecting them and being affected by them.

The puzzle of interacting actants can be seen as a network of heterogeneous agents, where links between agents are determined by the nature of interaction and its strength -- hence \textit{Actor Network} Theory. Furthermore, actants not only enter into networked associations, but themselves develop as networks -- of symbolically invested "things", "identities", relations -- capable of nesting within other diverse networks \citep{ritzer_actor_2005}. Actor Network Theory is about \textit{how} to study things by letting actants to emerge from the descriptions of interactions among diverse centres of influence within the ecology of symbolically invested "things" -- and then letting the actants "to have some room to express themselves" \citep{latour_using_2004}. Conversely to most social theories, \textit{ANT} does not posit any theory of the socio-technological world -- rather, it allows to study new, quickly changing and terribly fuzzy topics, where the framework of understanding them has to emerge during the very process of understanding. 

The significance of \textit{ANT} in Policy Scan and Technology Strategy design methodology is twofold. First, it gives a conceptual framework for dealing with organizational units, humans, laws, policy directions, technology domains, standards and solutions in terms of fluid \textit{actants} in the network of associations. Second, the concrete actants and their relations are socially constructed by identifying the nature of symbolic investment into them from the perspective of research questions (section \ref{research_questions}). The goal of the present research in terms of \textit{ANT} is to construct an actor network which would help us understand and navigate the interactions between social governance, expectations and problems on the one hand and technology developments in a specific context on the other.

\subsection{Boundary objects}\label{boundary-objects}

\textit{Boundary objects} are externalizations of ideas that are used to communicate and facilitate shared understandings across spatial, temporal, conceptual, or technological gaps \citep{wenger_communities_2010}. The concept was first introduced by sociologists S.L. Star and philosopher J.R. Griesemer in 1989 article \citep{star_institutional_1989}. Boundary object is a conceptual tool for managing a tension between the need to keep both diversity and coherence of concepts used in any work that involves innovating in an ecology of multiple actors and their perspectives. 

Since a boundary object emerges when multiple perspectives meet in a given ecology, it cannot happen prior to the actual interaction among actors holding these perspectives. It is therefore best to see boundary objects as actants in a Actor-Network which are furthermore \textit{created by the interaction with and of other actants}. A boundary object allows to consolidate differing perspectives as well as mediate between them without requiring for the actors in an ecology to converge to a single perspective, which is often impossible or impractical (albeit commonly preferred) in social and scientific world.

Boundary objects are objects which are both plastic enough to adapt to local needs and the constraints of the several parties employing them, yet robust enough to maintain a common identity across sites and perspectives. They may have different meanings for local groups but with common enough structure to enable them to be recognized and discussed among the groups. The creation and management of boundary objects is a key process in developing and managing coherence across intersecting social worlds \citep{star_institutional_1989} and perspectives.

\begin{figure}[H]
  \center
  \noindent\makebox[\linewidth]{\rule{\textwidth}{0.4pt}}
	\begin{subfigure}[c]{0.45\textwidth}
	  \noindent
      \includegraphics[width=1\textwidth]{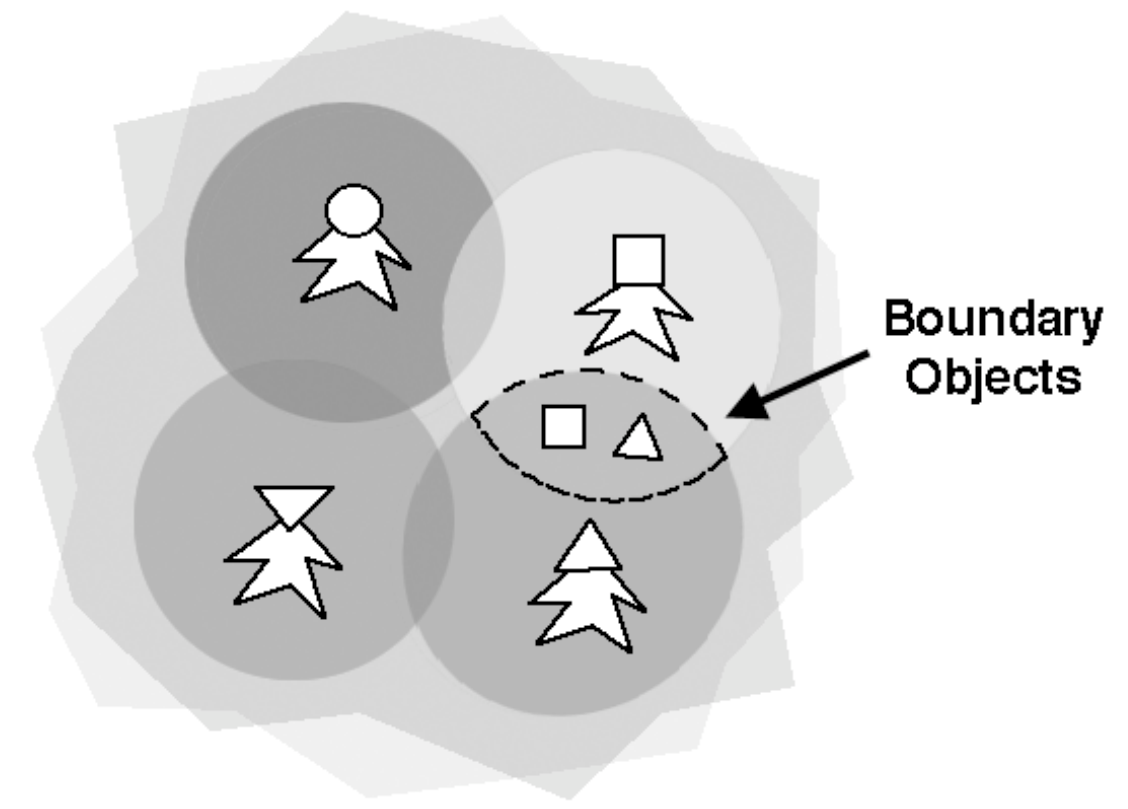}
      \caption{Boundary objects should be meaningful within the conceptual knowledge systems of at least two communities of practice. The meaning need not be the same—in fact, the differences in meaning are what lead to the creation of new knowledge. Adapted from \citet{wenger_communities_2010}.}
      \label{fig:bounary-objects-general}
    \end{subfigure}%
    %~ %add desired spacing between images, e. g. ~, \quad, \qquad, \hfill etc.
  \qquad
    \begin{subfigure}[c]{0.45\textwidth}
		\includegraphics[width=1\textwidth]{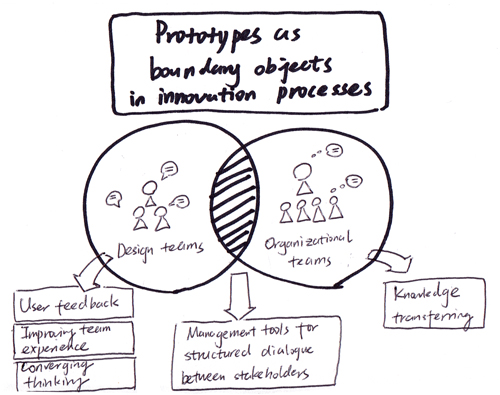}
        \caption{Prototypes as boundary objects in innovation process. Adapted from \citet{yin_boundary_2013}.}
        \label{fig:boundary-objects-in-innovation}
    \end{subfigure}%
    %~ %add desired spacing between images, e. g. ~, \quad, \qquad, \hfill etc.
\caption{Graphical description of boundary objects in interactions.}
    \label{fig:boundary-objects-illustrations}
\end{figure}

Creation and management of Actor-Network--embedded boundary objects of knowledge and practice is a way to approach and facilitate innovation \citep{akrich_key_2002-1} and new product development \citep{carlile_pragmatic_2002}, which we employ in this research. In such a network, boundary objects themselves become actants which influence and shape the interaction and development of new ideas, products and innovations.

\subsection{Change management in complex social systems}\label{change-management-in-complex-systems}

Change management in complex social systems was developed by co-authors of this paper \citep{veitas_system_2007} specifically for public policy design involving multiple stakeholders. It is a method for systemic management of projects, programmes, governance and policy initiatives which affect multiple interest groups with diverse and conflicting world-views and perspectives. It integrates the social system design with qualitative design methods of grounded theory construction for describing a way to maximally align the diverse interests of multiple stakeholders along the desired direction of social system development. 

\begin{figure}[H]
  \center
  \noindent\makebox[\linewidth]{\rule{\textwidth}{0.4pt}}
  \begin{subfigure}[c]{0.55\textwidth}
    \noindent
      \includegraphics[width=1\textwidth]{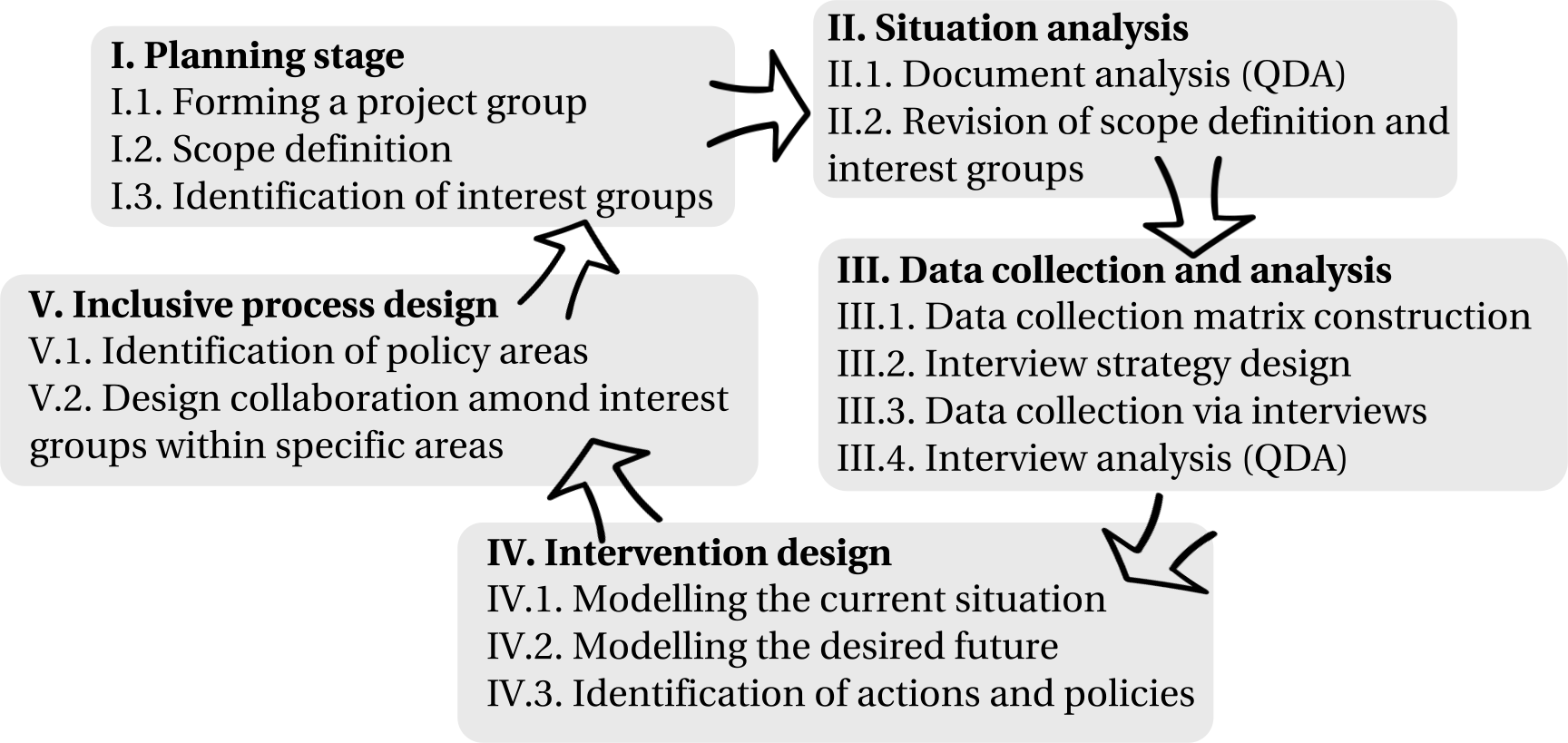}
      \caption{The schematic description of the process.}
      \label{fig:change_management_process}
    \end{subfigure}%
    %~ %add desired spacing between images, e. g. ~, \quad, \qquad, \hfill etc.
    \begin{subfigure}[c]{0.45\textwidth}
    \includegraphics[width=1\textwidth]{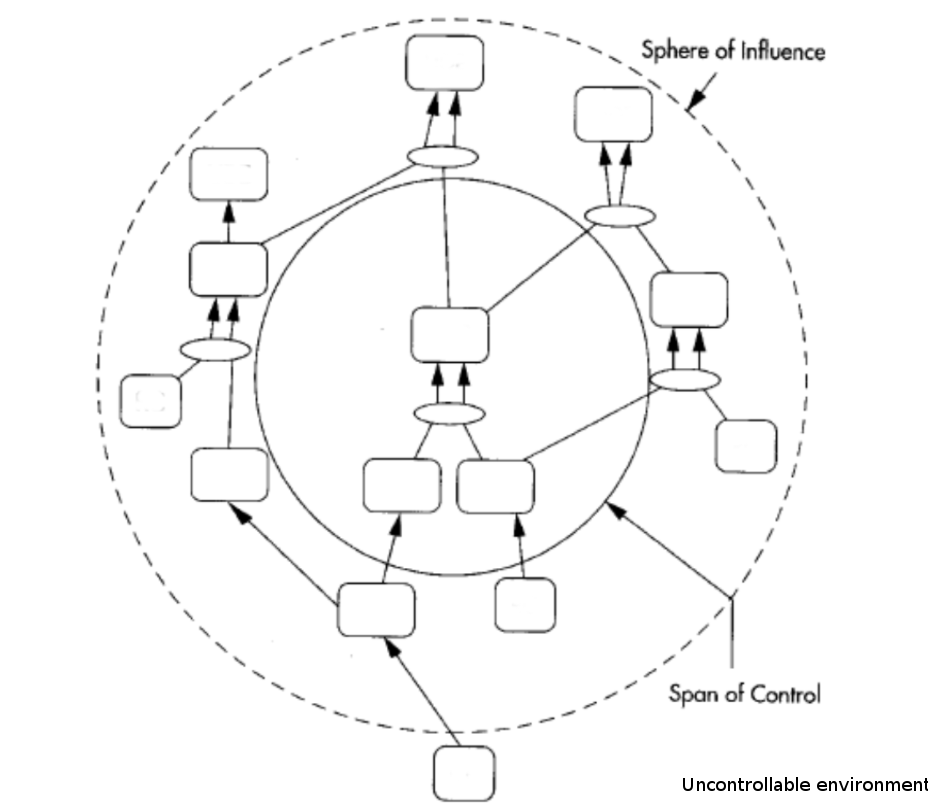}
    \caption{Systemic model of interactions among vested interests.}
    \label{fig:change_management_model}
    \end{subfigure}%
    %~ %add desired spacing between images, e. g. ~, \quad, \qquad, \hfill etc.
\caption{The method of change management for complex social system -- graphical depiction of the process and systemic model (adapted from \citep{veitas_system_2007}).}
    \label{fig:complex-system-change-methodology}
\end{figure}

Figure \ref{fig:complex-system-change-methodology} describes the process in a nutshell. First, the influence groups and their vested interests are identified through document analysis and open interviews with key representatives of a social system (Figure \ref{fig:change_management_process}). All relevant aspects of the complex system are then mapped into a systemic model which explicates the connections and interactions between the interests and through them -- groups of the social system which hold these interests (Figure \ref{fig:change_management_model}). Note that in Policy Scan and Technology Strategy design (Section \ref{policy-scan-methodology}), the technological artifacts themselves are considered interest groups of the system and "actants" of the network. The resulting network of stakeholders, their interests and interactions is then used for finding out win-win solutions for most or all stakeholders including, but not limited to the owner of the initiative. This way the social system of interest is guided towards the desired change.

The methodology for policy scan and technology strategy design introduced in the next section combines these conceptual approaches and methods in order to explicate regulation domains, current and future legislation, specific relevant statements and paragraphs in documents, research papers, standards, organizations, research groups, technical solutions, hardware and software components into an actor-network of boundary objects describing the context for developing a technology strategy or a specific technological solution in a chosen domain which mitigates concrete or developing policy requirements.

\section{Policy scan and technology strategy design methodology}\label{policy-scan-methodology}

\begin{figure}[H]
  \center
  \noindent\makebox[\linewidth]{\rule{\textwidth}{0.4pt}}
  \includegraphics[width=0.9\textwidth]{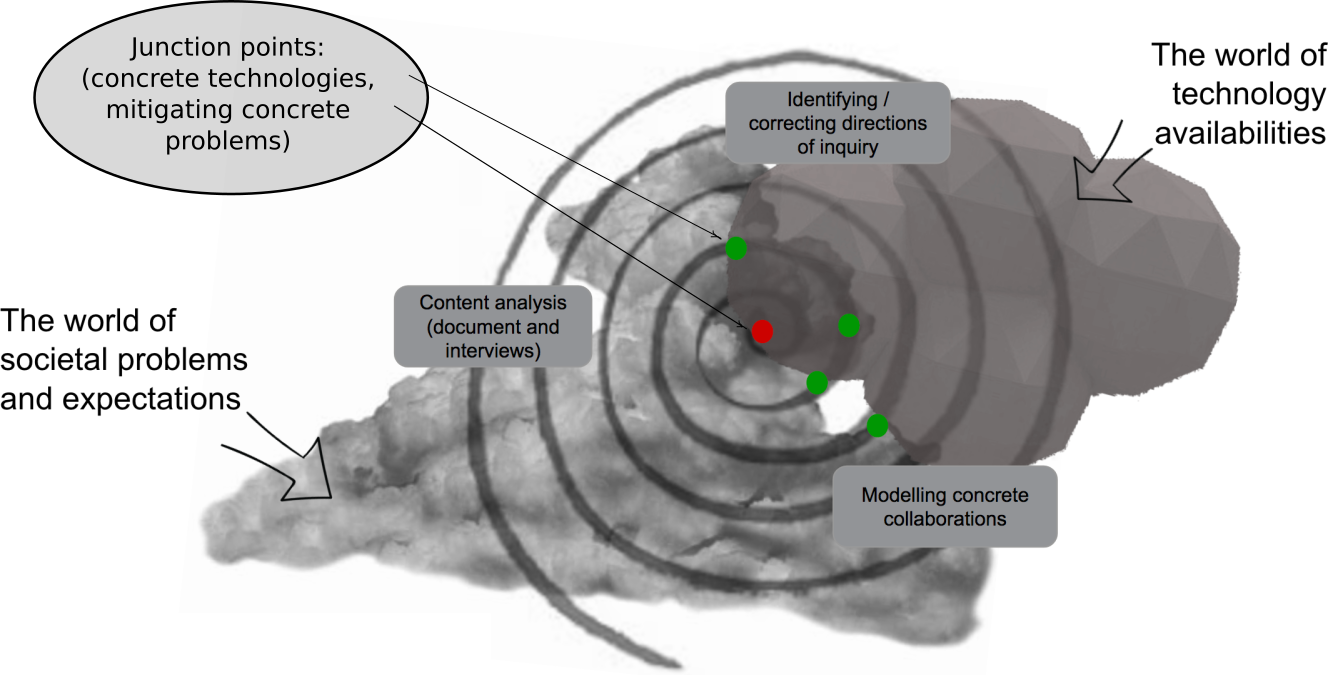}
  \caption{The iterative process of policy scan and technology strategy design as a methodology of relating the world of societal problems and expectation with the world of technology via their junction points -- concrete problems that can be mitigated with concrete technology solutions.}
  \label{fig:relating-two-worlds}
\end{figure}

Policy Scan and Technology Strategy design methodology is aimed at consolidating connections between the "application pull" of implicit societal problems and expectations with the "technology push" of technological availabilities and developments in terms of concrete solutions mitigating concrete problems. Such connections span several domains of activity, including technology building blocks and platforms, hardware and software design, system specifications, applications, business models, policy requirements and societal expectations. Consolidated connections act as junctions between the world of societal problems and expectations and the world of technological availabilities and drive a long term innovation road-map of a company or organization that engages in designing them (Figure \ref{fig:relating-two-worlds}).

The challenge lies in identifying and developing the junctions between two highly uncertain and dynamic domains (of societal problems and technological availabilities). Policy Scan and Technology Strategy design approaches this challenge by creating the ground for interaction from which the junction points consolidate through the iterative process of negotiating an matching requirements, expectations and histories of the respective domains. 

\begin{figure}[H]
  \center
  \noindent\makebox[\linewidth]{\rule{\textwidth}{0.4pt}}
  \includegraphics[width=0.7\textwidth]{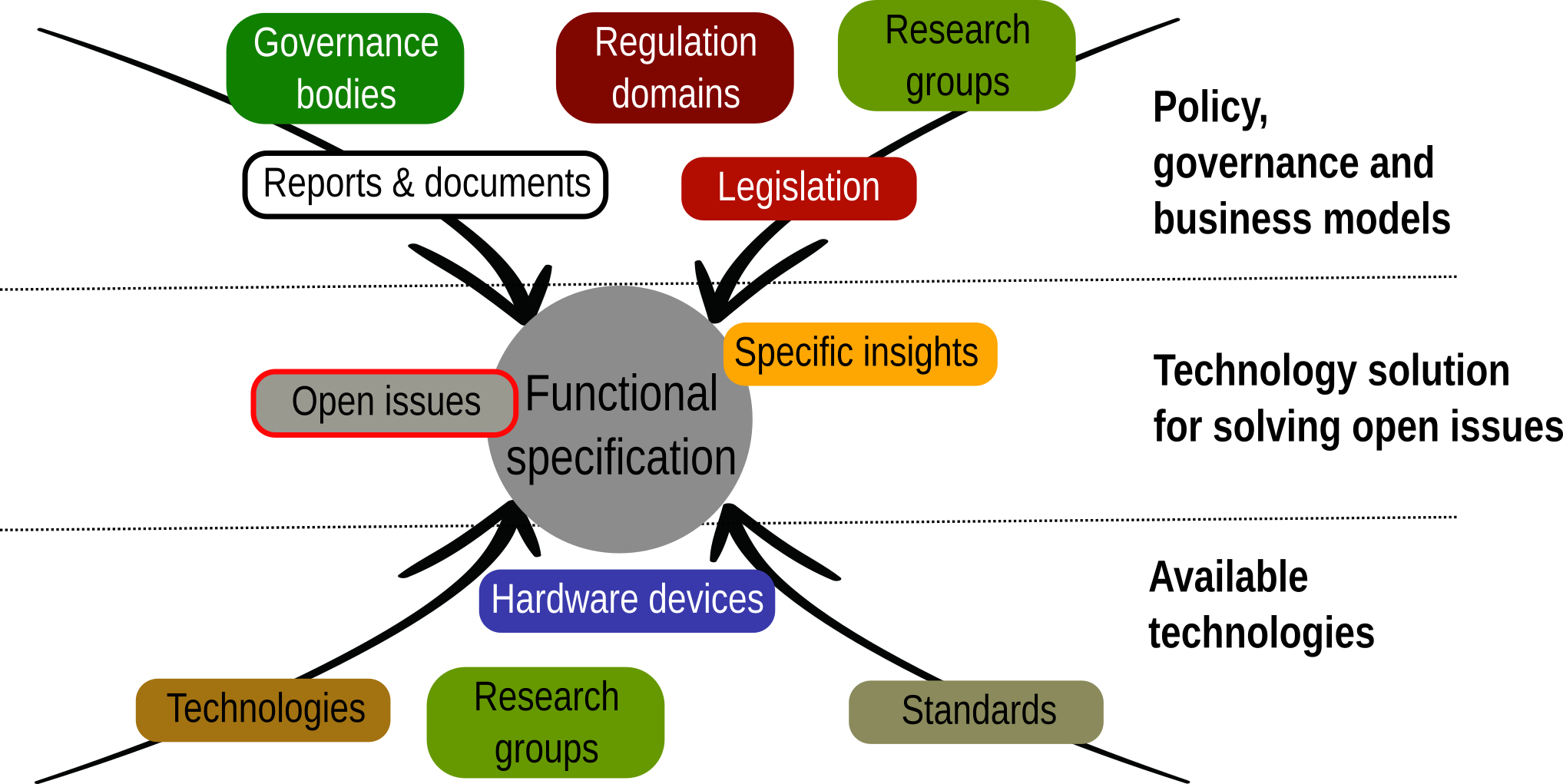}
  \caption{Double funnel of policy scan and technology strategy design.}
  \label{fig:integration-funnel}
\end{figure}

The accelerating disruptive technological development imposes new requirements for the governance models, regulatory regimes and, most importantly, the legislation creation processes which have to become much more flexible and adaptable. Legislation and policies cannot any more be in development for years and stay unchanged for decades after enforcement. The legislative environment should be ready before the technology is deployed and this creates the situation when policy makers have to produce a "future-proof" legislation for regulating not-yet-fully-known but already disruptive technology ecosystems and business models \citep[p. 26]{dg_grow_gear2030_2017}. At the same time, technology should be flexible enough to adapt to constant changes in legislation. The challenge is to fashion regulations that allow enough flexibility in the future but also meet the safety, security and privacy concerns of the public and policy makers \citep{ranft_freeing_2016}. The process that addresses this challenge cannot be seen as 'top-down' or 'bottom-up' -- it looks much more like a constant 'negotiation' between developing regulatory and technological solutions, which requires both to be flexible and adaptable. Policy Scan and Technology Strategy design methodology offers to account for these circular interdependencies via identification of specific technological solutions which would serve as connecting tissue, areas of junction and boundary objects (see Figure \ref{fig:relating-two-worlds}) between policy, governance, business modelling domain and the domain of available technologies (see Figure \ref{fig:integration-funnel}). Note, however, that the solution is never chosen \textit{a priori} to the process but emerges during and as a result of it.  

In spite of the fluidity and iterative nature of the process there still is a value in describing it in terms of stages. It is important however to always have in mind, that while it is useful to think in terms of separate stages, they are not clearly chronologically ordered -- usually the stages of the policy scan and technology strategy design overlap in terms of time and content and are constantly iteratively visited and revisited during the process (Figure \ref{fig:relating-two-worlds}). A better way to approach this process as a \textit{design inquiry} -- an operation to create something useful and usable in a specific domain and situation -- as contrasted to proving what is generally best or optimal.

\subsection{Stage 0: Content analysis (documents and interviews)}

The process starts from a literature review (reading, analyzing and open-coding reports, newspapers, web articles, research papers, etc.) in a relevant area. Content analysis is performed in a manner similar to the construction of a grounded theory (see Section \ref{grounded-theory}) and is carried throughout the whole time of the project, overlapping with all the other stages. The manner in which the stage is carried out at each iteration changes with the maturity of the design inquiry process. At the start it is mostly inductive and based on the \textit{open coding} of data -- an analytic process by which categories are created and attached to the observed data without considering any a priory explicit theoretical framework. With the emergence of recurring ideas, concepts and categories they are integrated into a conceptual framework via \textit{axial coding} -- a process of relating categories identified via open coding to each other through iterations of inductive and deductive thinking. Finally, when a network of concepts, ideas, technologies and artifacts starts to take shape, the process proceeds to deductive verification of the constructed model via \textit{selective coding} -- a process of developing a single storyline of why categories are related to each other -- i.e. selecting core categories and relating other concepts to them \citep{bohm_theoreticai_2004}. The obtained data is used as an input for Stage 1. Note, that the efficiency of data collection and analysis stage requires heavy use of Qualitative Data Analysis (QDA) as well as business modeling methodologies and tools.

\subsection{Stage 1: Identifying / correcting directions of inquiry}\label{stage-1}

As soon as the open codes and concepts identified during Stage 0 start to connect and consolidate into larger associative networks, it makes sense to start outlining the scope of the inquiry by differentiating the relevant avenues from the ones that are most probably out of scope for the initiated design inquiry. At the same time, a list of people (researchers, policy makers, technology officers and business managers) start to emerge -- we shall refer to them as the \textit{gatekeepers} of the design inquiry. Involving gatekeepers into the design process by interviewing them and discussing the alternative avenues to be pursued, verifies initial ideas and concepts obtained from literature review as well as further guides the design inquiry towards concrete solutions. Simultaneously and related to that, the topical areas of interest start to consolidate and evolve into the information requirements for advancing the design further. Usually, additional documents and research domains are identified, reviewed and analysed by revisiting Stage 0 activities -- so there is a positive feedback loop between Stage 0 and Stage 1 (and, as it will become apparent, between all subsequent stages). The approach that keeps this positive feedback loops from exploding in terms of associations, relations and possible avenues is the disciplined focus on the pragmatic direction of the inquiry -- i.e. designing a concrete technology strategy and solution which necessary have real world constraints. Note again, that the scope definition is never complete but constantly being verified, updated and further specified thus funneling the data and information towards the direction of the design inquiry.

The aspects that have to be taken into account when designing a scope definition are described below. Note, that they are described in terms of the chosen application domain -- smart mobility and autonomous driving technologies which served as a use case for designing the Policy Scan and Technology Strategy methodology. In Section \ref{identifying-junction-points} a further description of how the methodology was applied to the application domain is provided.

\begin{description}

\item[Target technology domain.] The identification and consolidation of a technology domain with a strong potential to mitigate societal problems with technology solutions is the starting (and focal) point of the process. Smart mobility and autonomous driving is a quickly developing market, driving the research and commercial demand for smart sensors, hardware, as well as a myriad of digital technologies (see Section \ref{identifying-junction-points} for the explanation of how this application domain was identified using the methodology). Furthermore, it is a part of an even larger ecosystem of Internet of Things. The pragmatic way to navigate this emerging landscape from technology strategy design perspective is to simultaneously consider existing solutions, research expertise of the owner of design inquiry, dynamics of the whole intelligent transportation systems domain, the readiness of specific technologies to market deployment and, last but not least, the perceived need of these technologies within the emerging ecosystem.

\item[Markets] are strongly related to the directions of technological development and target domains, but are not always the same. For example autonomous mobility has many faces - i.e. truck platoons, private cars, first/last mile solutions, autonomous taxi fleets, collaborative driving, etc. Technologies that are needed for them can be quite different. The consideration of market segments when designing technology strategy may help to forecast which technology has good chances to succeed in which segment and to navigate accordingly.

\item[Strategic considerations] can be broken into two aspects:
  \begin{enumerate}
   \item Areas of interest, sphere of influence and span of control of the owner of design inquiry in terms of how much and how far the influence on the environment and other actants in the network can be exerted. Obviously, it is not possible for a single organization to target all markets, all technologies and all usages at the same time (especially considering the background in certain technologies of the owner of design inquiry), therefore the priorities about where to put money, efforts and(or) lobbying power should be considered in terms of pragmatic limits of span of control of the owner of the design inquiry (see Figure \ref{fig:change_management_model});
   \item Desirable \& undesirable effects. After identifying domains of interest, it is possible to start considering what is desired and possible to achieve with respect to the current situation in a chosen application domain -- i.e. what are the desirable and undesirable effects of the technology strategy actions;
  \end{enumerate}

\item[Legal and policy aspects.] First, legal and policy related constraints can be reasonably estimated only for specific technologies and their usages. For example, legal requirements for autonomous vehicles depend very much on whether it is a truck operating on a highway, closed logistic center or a passenger car in an inhabitable area. Second, the legal framework and regulatory requirements level is not uniformly developed for different technologies. E.g. existing technical standards in terms of security can be identified and applied for the sensor development; yet the legislation for the autonomous driving, as well as privacy / security of IoT devices is still evolving, albeit being guided by general data privacy, provenance and security policies. While some of the existing regulations can be applied, it is not clear how much they are going to change to accommodate the autonomous driving technology. Finally,  legislation and/or policy is currently being formulated at national, European or international levels, therefore certain lobbying activities could be considered in order to influence legislative actions in a favourable way -- the scope definition may inform well of potential directions of such activities.

\end{description}

\subsection{Stage 2: Modeling the solution}

\begin{figure}[H]
  \center
  \noindent\makebox[\linewidth]{\rule{\textwidth}{0.4pt}}
  \includegraphics[width=0.7\textwidth]{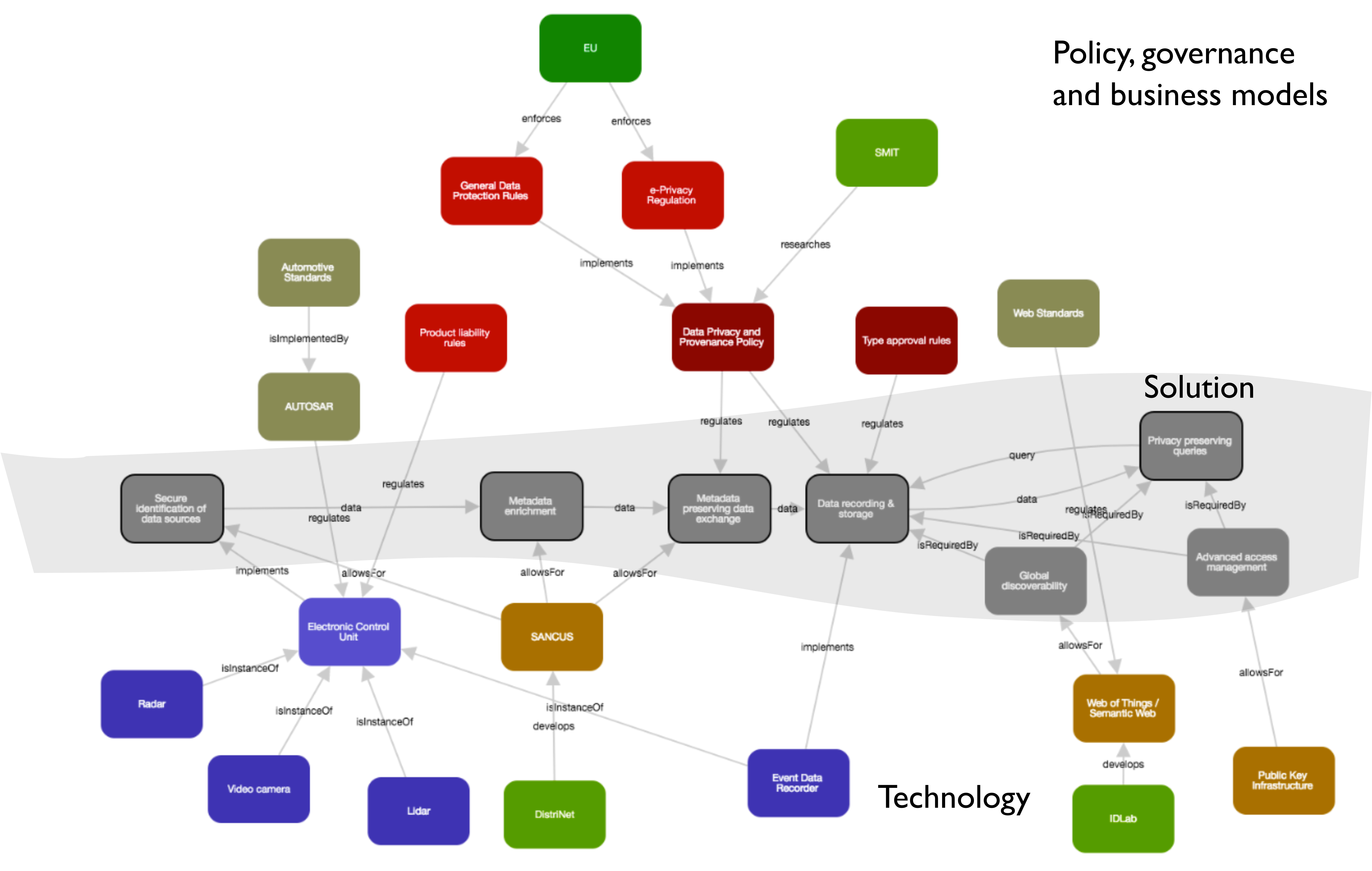}
  \caption{The network of a policy requirements and technology availabilities with respect to a specific technological solution}
  \label{fig:black-box-network}
\end{figure}

The scope and landscape of the inquiry mapped during Stage 1 allow to concretely identify one or more specific technology solutions to be developed further along the overall technology strategy in a given domain. An important principle of the policy scan and technology strategy design methodology is that it aims not only at the description of general domain of target development, but also identifies very concrete solutions that can be further researched or developed towards market deployment. During Stage 2 these solutions are identified and further specified in terms of an integrated network of policy requirements and technology availabilities (Figure \ref{fig:black-box-network}).

Note the close relation between the Figure \ref{fig:black-box-network} and the double funnel depicted in Figure \ref{fig:integration-funnel}: the network of policy requirements and technology availabilities is obtained by specifying all aspects of the the double funnel with respect to a specific solution. This means that out of all documents, research groups, technology standards and other categories, reviewed during Stage 0, only the relevant knowledge is extracted and included into the network. For example, if the double funnel contains the consideration for data privacy requirements (as is in the case of chosen application domain of autonomous driving), the network of specific solution will contain concrete clauses and their interpretations from relevant legislation (e.g. General Data Protection Regulation). Furthermore, since in the quickly changing technology domains for which the methodology is created, the knowledge becomes obsolete soon, the network includes statements and references to expected future developments (e.g. upcoming opinions of existing working groups regarding relevant legislation). Last but definitely not least, the additional literature review and very targeted document analysis is usually needed for developing the network. Therefore, as already mentioned above, Stage 0, 1 and 2 are organized in an iterative loop which allows for the solution to become more specified with each iteration (see Figure \ref{fig:relating-two-worlds}).

\section{Identifying junctions between societal problems and technology worlds}\label{identifying-junction-points}

As we have pointed out earlier, the research into formulating the methodology for connecting the worlds of societal problems and technology was carried out within specific application domain of smart mobility and autonomous driving -- as operating within a specific context is a major methodological principle of Policy Scan and Technology Strategy Design.

The initial motivation for engaging into research was twofold: 
\begin{inlinelist}
  \item to develop a methodology for addressing specific types of "ill-defined" problems in terms of integrating technological development and societal expectations -- which resulted in Policy Scan and Technology Strategy Design methodology;
  \item to demonstrate the application of methodology in a specific context by developing a proposal for a concrete technological solution for mitigating policy concerns in a chosen application domain -- which resulted in the proposal for Data Storage System / Event Data Recorder of Autonomous Driving.
\end{inlinelist}
In this section we present and explain in detail how the concrete societal problems were identified and technology solutions consolidated while developing and simultaneously applying the Policy Scan and Technology Strategy Design methodology as a grounded theory process. The actual technology solution and related context is the subject of the second paper documenting the results of research \citep{veitas_-vehicle_2018}.

The rationale for choosing the application domain for developing methodology of relating the world of social expectations and technological availabilities was based on the observation of many uncertainties of integrating autonomous robot and AI technologies into the social fabric. The sales of robot technologies are currently increasing in double digits every year. Robots are increasingly becoming more autonomous and permeating almost every aspect of human life, which starts to influence the established regulations and disrupt social governance. Such dynamically developing domains with high uncertainty is precisely the areas in which Policy Scan and Technology Strategy Design provides most value as it consolidates a huge informational context into actionable innovation road-map for both companies and governments. 

\subsection{AI policy compliance}

Therefore, the first step towards the junction points between societal expectations and technology availabilities was the choice of the "AI policy compliance" application domain, concerned with how autonomous robot and AI technologies are going to be integrated into social governance structures, legal frameworks and regulations. While this domain is immensely broad, some governments, namely European Union, are starting to draft proposals for regulating it.

European Parliament's resolution "Civil Law Rules on Robotics" of February 2016 \citep{european_parliament_resolution_2017} is considered one of the most progressive and forward looking government initiatives in the world regarding status of autonomous robots in society. The document is not binding and does not impose any regulations, but provides well thought, rich and clear statements of challenges of development and social acceptance of robotic technologies as well as proposals how to approach them. The resolution considers many long and medium term aspects, including:

\begin{enumerate}
  \item there is a need for a generally accepted legal definition of robot and AI that is flexible and is not hindering innovation, yet allows for efficient legal treatment;
  \item in the long term, the current trend in robotics technologies leans towards developing autonomous machines with the capacity to be trained and make decisions independently (i.e. without human intervention) holds not only economic advantages, but also variety of concerns regarding direct and indirect effects on society as a whole;
  \item that machine learning and AI offers enormous economic and innovative benefits by vastly improving the ability to analyze data, while also raising challenges to ensure non-discrimination, due process, transparency and understandability in decision-making processes;
  \item a requirement that those involved in the development and commercialization of AI applications and robots build in security and ethics considerations at the development phase, accepting legal liability for the quality of the technology they produce;
  \item there is a possibility that in the long-term, AI could surpass human intellectual capacity, an event that potentially would have a disruptive effect on the foundations of socio-economic organization of human society and legal systems;
  \item it is important to bear in mind that developing robotics in the face of increasing divisions of society with shrinking middle class and increasing capital concentrations may lead to a high concentration of wealth and influence in the hands of minority affecting the democratic principles of social governance to the point of no return;
  \item AI and robot technologies can and should be designed in a way that they preserve dignity, autonomy and self-determination of the individual;
  \item the current legal framework would not be sufficient to cover the damage caused by the new generation of robots, insofar as they can be equipped with adaptive and learning abilities entailing a certain degree of unpredictability in their behavior, since those robots would autonomously learn from their own variable experience and interact with their environment in a unique an unforeseeable manner;
\end{enumerate}

As a general principle, European Parliament proposes to follow "gradualist, pragmatic and cautions approach" with regard to future initiatives on robotics and AI so that not to stifle innovation yet still considering the challenges involved. In terms of research and innovation direction, the EP believes that interoperability among systems, devices, robots and cloud services, based on the security and privacy by design (i.e. European data economy\footnote{\href{https://ec.europa.eu/digital-single-market/en/policies/building-european-data-economy}{https://ec.europa.eu/digital\-single\-market/en/policies/building\-european\-data\-economy}}) is essential for real time data flows enabling robots and AI to become more flexible and autonomous, and that it will require large scale digital infrastructure providing ubiquitous connectivity.

Most importantly for our research, from the ethical perspective, the resolution highlights and proposes \textit{the principle of transparency}, which is made of three components:

\begin{enumerate}
  \item it should always be possible to supply the rationale behind any decision
taken with the aid of AI that can have a substantive impact on one or more persons’
lives;
  \item it must always be possible to reduce the AI system’s computations to a form comprehensible by humans;
  \item advanced robots should be equipped with a "black box" which records data on every transaction carried out by the machine, including the logic that contributed to its decisions.
\end{enumerate}

From the governance perspective, European Parliament asks the European Commission to consider the designation of a \textit{European Agency for Robotics and Artificial Intelligence} in order to provide technical, ethical and regulatory expertise needed to support public authorities. Last but not least, it calls for the analysis and consideration of different legal solutions, including creating a specific legal status applicable for advanced autonomous robots in the long run, possibly creating and applying the civil status of electronic personality to cases where robots make autonomous decisions and interact with third parties and environment independently.

\subsection{Autonomous vehicles and smart mobility}

One of the fastest developing areas in the domain of autonomous robotics and AI with a promise of a large scale deployment in short or medium term is self-driving technologies and smart mobility systems based on them. Note, that an autonomous vehicle fits precisely the definition of a general autonomous robot (as used in European Parliament's resolution) in that it learns from environment and makes independent decisions that affect third parties. Moreover, participation of autonomous vehicles in transport systems raises immediate legal and liability challenges of the sort pointed out in "Civil Law Rules on Robotics" by EP\citep{european_parliament_resolution_2017}.

First truly autonomous vehicles appeared in 1980s in the form of research projects by major universities and car manufacturers. In 1990s and 2000s large scale research projects in Europe and US have demonstrated self-driving cars that were able to travel thousands of kilometers in mixed traffic conditions on public roads essentially performing all driving functions. The 2010s marked the accelerating commercial interest in autonomous driving technologies, with major companies (first technology companies and then established vehicle manufacturers) starting projects for developing prototypes of autonomous or semi-autonomous vehicles with clear market deployment prospects. As of 2018, all major car manufacturers as well as some technology companies have an autonomous vehicle in development and(or) already sell cars with limited self-driving capabilities (e.g. park-assist, lane change functionality, etc.). At the same time, governments, cities and traffic authorities have realized both the potential offered by the integration of self-driving technologies into transport infrastructure and public transportation systems as well as related legal, regulatory and infrastructural challenges. At the time of writing, major regulatory initiatives for adapting traffic rules, international agreements, product and civil liability rules and more are well under way at the level of United Nations, European Union, European member states, US federal and state levels.

\begin{figure}[h]
  \center
  \noindent\makebox[\linewidth]{\rule{\textwidth}{0.4pt}}
  \includegraphics[width=0.7\textwidth]{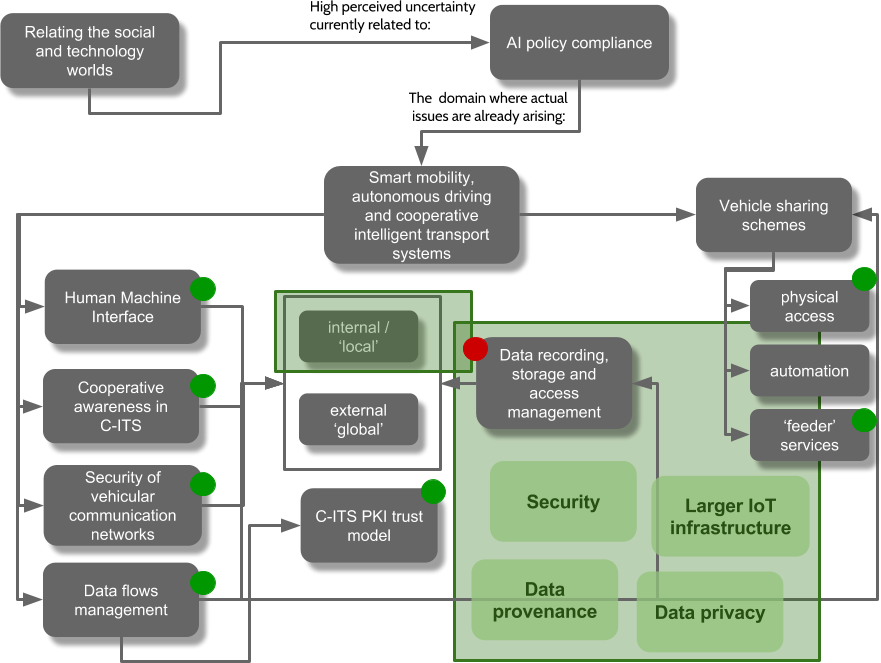}
  \caption{Choices and considerations illustrating how the double funnel of Figure \ref{fig:integration-funnel} led to the identification of in-vehicle data recording, storage and access management as a junction point between the world of societal problems and technology availabilities in the domain of autonomous driving and intelligent transport systems using Policy Scan and Technology Strategy Design methodology. Green dots illustrate potential junction points between the social problems and technology worlds. The red dot is the chosen concrete societal problem and technology solution further discussed in \citep{veitas_-vehicle_2018} (see also Figure \ref{fig:relating-two-worlds} for illustration of junction points between worlds of social problems and technology availabilities).}
  \label{fig:decision-tree}
\end{figure}

Considering maturity, proximity to market deployment, fast development and the amount of financial, research and regulatory resources currently being invested into the autonomous driving and smart mobility systems, they are a perfect candidate for developing and applying Policy Scan and Technology Strategy design methodology for creating junctions between social and technology worlds. Most importantly, the methodological principle of exercising and researching actual interaction between the worlds is readily available as both self-driving technologies and societal expectations towards them are already well consolidated and represented in the public debate. Therefore the "AI policy compliance" research direction was further funneled towards autonomous driving technologies and smart mobility systems for the identification of concrete societal problems and technological solutions (see Figure \ref{fig:decision-tree}). 

\subsection{Potential junction areas}

The domain of autonomous vehicles and smart mobility is complex, multifaceted and encompasses many societal problems, expectations and requirements that can potentially be mitigated by different types of technological as well as governance solutions. The main potential junction areas considered in the course of research are:

\begin{description}
  \item[Human Machine Interface] (HMI), which is a very important aspect in automated driving scenarios since autonomous vehicles will have to interact (present integrated traffic information and ask for feedback or decision) with a driver (internal HMI) as well as other road users (external HMI). One of the most pressing issues is the one related to transferring the control of a vehicle between human and machine drivers in hybrid or semi-autonomous driving scenarios. HMI rules will have to be drafted based on user research, security considerations, technical capabilities, etc. and most probably included into type-approval rules of autonomous cars.
  \item[Cooperative awareness] refers to automated driving scenarios where local driving decisions of a vehicle are made based on the integration of data and intelligence from surrounding vehicles, road-side units and potentially other traffic participants. It is based on the realization that real-time interaction between vehicles and road-side units allows for more efficient traffic management both locally and globally. E.g. sharing information or even raw sensor data feeds between vehicles and stationary road-side units for allowing collective decisions may drastically improve the security of each traffic participant as well as efficiency of the whole system - which is therefore called a \textit{Collaborative Intelligent Transportation System (C-ITS)}. 
  \item[Security of vehicular communication networks.] Current problems with security of vehicular communication networks, which have been subject to a growing number of attacks that put the safety of passengers at risk and cost huge losses for automotive manufacturers \cite{van_bulck_vulcan:_2017} increase exponentially when considering automated driving and collaborative intelligent transport systems, where it will not be possible to completely close the in-vehicle networks from out-vehicle communications. Ensuring security of the collaborative intelligent transportation system is the major aspect of developing these systems in the first place.
  \item[Public Key Infrastructure (PKI) trust model.] PKI is comprised of a set of technologies that facilitate secure electronic transfer of information over an insecure network that provides roles, policies and procedures to manage digital identities of information senders and receivers. Integration of solutions of automotive component authentication and software isolation \cite{van_bulck_vulcan:_2017} with C-ITS PKI system and trust model is a promising avenue for research and product development, as these solutions will largely determine the usability and scale / speed of the deployment of the whole system.
  \item[Data flows management.] Large scale deployment of the collaborative intelligent transportation system technologies will entail the collection, exchange and processing of unprecedented amounts of data, including, but not limited to broadcasts of messages about vehicles' location, sensor data, awareness of its surroundings, and more. Part of these messages will contain private information that will have to be protected according to EU and international regulations.
  \item[Data recording, storage and access management.] One of the aspects of data flows management is that at least part of the internal (in-vehicle) and external (between vehicles) messages will have to be recorded and stored for variable periods of time not only in order to ensure the efficient operation of the C-ITS as a whole, but also for on-line and off-line analysis of operation and traffic events (e.g. accidents). Considering many issues with vehicular communication networks security, data privacy and provenance requirements, the robust access management system is instrumental to the very possibility of the deployment of the C-ITS. On the one hand data recording is a requirement for semi-autonomous systems (very much like "black boxes" of the airplanes). On the other hand, data recording devices are in line with the European Parliament's recommendations on "Civil Law Rules on Robotics"\citep{european_parliament_resolution_2017}.
  \item[Vehicle sharing schemes.] The magnitude of the prospective C-ITS economic and social benefits is crucially dependent not only on automation and cooperation between individual vehicles, but -- even more so -- on effective vehicle sharing schemes. It is estimated, that currently, an individual vehicle is used on average about 5\% of the time, while the rest of the time it occupies a parking space. Vehicle sharing schemes would ensure much higher rate of utilization. Further, automated driving may increase number vehicles on the road due to their convenience, which would increase the stress on already overloaded road infrastructure in many cities. The only perceived way to counterweight such dynamics is the implementation of efficient and acceptable by users sharing schemes which is one of the main policy directions at cities level. Vehicle sharing will posit additional challenges as well as research opportunities for data privacy, provenance and security (e.g. key-less car sharing schemes \citep{symeonidis_sepcar:_2017}).
\end{description}

\subsection{In-vehicle data recording, storage and access management}

Out of the potential junction points listed above we have selected the \textit{in-vehicle data recording, storage and access management} for further consolidation using Policy Scan and Technology Strategy Design methodology (see Section \ref{stage-1} "Stage 1: identifying / correcting directions of inquiry" for the explanation of aspects that were taken into consideration).

The in-vehicle data recording, storage and access management technology (i.e. device) is the closest to the actual implementation and market deployment in terms of additional research and development needed, readiness of regulatory requirements and already existing technologies in vehicles as well as other industries. Concretely, in-vehicle data recording and storage requirements are going to be included into type-approval rules of autonomous vehicles. Taking into account the large scale plans (at least in European Union) to roll out the collaborative intelligent transportation system beginning in year 2019/2020 there is a clear opportunity to develop and introduce this technology into the market. Furthermore, automotive in-vehicle data recording, storage an access management solutions can be seen as predecessors of the "black box" technologies for recording transactions and logic of advanced autonomous robots in general \citep{european_parliament_resolution_2017} -- providing a potential innovation pipeline serving one of the fastest growing markets in the world. At the same time, they are direct successor of the Event Recorder Devices (EDRs) currently being installed in all vehicles.

The in-vehicle data recording, storage and access management technology is therefore a clear junction between the societal problems and expectations and technology availabilities. The target societal problem in terms of autonomous vehicles and advanced robots in general is to be able to understand and retrospectively analyze their independent decisions as well as explain the rationale behind them, especially when these decisions affect third parties. In terms of technological availabilities, there are no fundamental research problems preventing the development and implementation of data recording and storage devices integrated into automotive CAN buses (see the follow up article \citep{veitas_-vehicle_2018} for an in-depth discussion of the technology and the surrounded context).

Notwithstanding all above, that biggest potential hurdle for developing and deploying the technology is the need to cooperate closely with vehicle manufacturers, since the technology will have to be integrated. Due to quite conservative and closed nature of automotive manufacturers' technology development process this may be a serious impediment not only for the potential market deployment of the in-vehicle data recording technology, but also for the large scale deployment of the C-ITS itself, which will legally require this technology. From the few contacts with vehicle manufacturers during this research we have formed an impression, that while the issue of in-vehicle data recording is being known, the in-house development efforts are rather limited. We strongly think that vehicle manufacturers should put more attention and possibly seek for collaborations for implementing and integrating into the vehicles the secure data recording and storage technology if they are developing and planning market deployment of a self-driving cars.

\section{Conclusion}

The accelerating technological development in variety of domains, including robotics, Artificial Intelligence, biotechnologies and genetics, nano-technologies, space travel, material science, alternative power generation and many more is already disrupting communities, societies and governance structures. The impact of technology to human societies has been known from the dawn of civilization, but the mere speed and magnitude of modern technology development is unprecedentedly changing the established way of life (and work). While the disruption seems to be inevitable, the actual direction of change is not predetermined -- it may radically improve the quality of life of all citizens, increase the availability to education, travel, communication, culture and business; or it may not stand to the global challenges like global warming, increasing income inequality, the fusion of governance and global capitalism structures and many more. The increasing dynamics makes the society as a whole as well as individual communities and cities more fragile in many ways, yet trying to contain the change may create even more unbearable tensions for the global socio-technological system. Therefore, the smooth integration of developing technologies into (rather conservative) social and business structures by providing a platform for negotiation between the world of social problems and expectations and the world of technological availabilities is of utmost importance.

Policy Scan and Technology Strategy Design methodology was developed precisely for finding specific junction points between the two worlds. The outcome of a concrete design initiative following the methodology is a clear and actionable innovation road-map or a specific innovation that mitigates specific problems. The methodology is therefore best applied from the perspective of technology companies which are interested in development and market deployment of socially acceptable and sustainable large scale technological solutions.

Te developed methodology is first of all a design initiative, which by definition requires establishing and facilitating actual interaction between the parties of societal world and technology world. Such design initiative cannot be a theoretical exercise but has to be developed and applied in real settings which touch actual problems and maximize the real interaction between technology developers, policy makers and citizens.  

In this research, we have developed the Policy Scan and Technology Strategy Design methodology by applying it to the problematics of integrating self-driving technologies to the transport systems. It is a use case that showcases both the abilities and power of the methodology to come up with specific technological solutions as well as helped to formulate the methodology itself. For the explanation of the actual technological solution that mitigates specific societal problems in a chosen domain -- \textit{in-vehicle data recording, storage, and access management solution} --, see the follow up paper describing results of the research project \citep{veitas_-vehicle_2018}.

The results of this research are broadly applicable beyond the provided use case which requires good understanding and fluid adaptation of the conceptual foundations of the methodology to every specific case of developing junctions between social world and technology world.

% === BIBLIOGRAPHY === 

\printbibliography[heading=bibintoc]
\end{document}